# Bayesian models are better than frequentist models in identifying differences in small datasets comprising phonetic data

Georgios P. Georgiou[1,2]

[1]Department of Languages and Literature, University of Nicosia, Nicosia, Cyprus

[2]Director of the Phonetic Lab

**Abstract**

While many studies have previously conducted direct comparisons between results obtained from frequentist and Bayesian models, our research introduces a novel perspective by examining these models in the context of a small dataset comprising phonetic data. Specifically, we employed mixed-effects models and Bayesian regression models to explore differences between monolingual and bilingual populations in the acoustic values of produced vowels. Our findings revealed that Bayesian hypothesis testing exhibited superior accuracy in identifying evidence for differences compared to the posthoc test, which tended to underestimate the existence of such differences. These results align with a substantial body of previous research highlighting the advantages of Bayesian over frequentist models, thereby emphasizing the need for methodological reform. In conclusion, our study supports the assertion that Bayesian models are more suitable for investigating differences in small datasets of phonetic and/or linguistic data, suggesting that researchers in these fields may find greater reliability in utilizing such models for their analyses.

*Keywords*: mixed-effects models, Bayesian regression models, phonetics; posthoc test; hypothesis testing

## 1 Introduction

Statistics is an integral part of quantitative research. Experimental linguistic research exploits quantitative methods, which require the use of statistical modelling. The necessity to elicit multiple data points from each participant corresponding to various levels of the linguistic variable under consideration led researchers to move from traditional ANOVA measures to more sophisticated tests that incorporate repeated measures. To this end, in the last few decades, the vast majority of phonetic research employs linear mixed-effects models for data analysis (Chen 2008; Escudero et al., 2012; Georgiou, 2022a; Georgiou & Dimitriou, 2023; Georgiou, 2023; Mills et al., 2022; Reinisch & Sjerps, 2013; Simonet, 2011). These models usually include fixed and random factors. As research methodologies advance, emerging models strive to provide more precise data analysis. Among the alternatives to frequentist statistics, Bayesian statistics have gained recent popularity. The application of such models in phonetic science is very limited (Escudero et al., 2022; Georgiou, 2022b).

Bayesian modelling presents several advantages. First, there is flexibility in the specification of the model parameters, which is difficult to do in frequentist statistics (Nicenboim &Vasishth, 2016). For example, the model can be tuned based on previous beliefs (Vasishth et al., 2018). This can be achieved by establishing specific parameters, such as degrees of freedom, means, standard deviations, and others, derived from previously observed data. After examining the data, the prior information is revised to form the posterior distribution, which is then utilized for making inferences (Puga et al., 2015). Second, the uncertainty regarding the magnitude of an effect can be quantified by identifying the range of plausible credible values that capture the potential effect (Vasishth et al., 2018). The Bayesian confidence interval (CI), which is known as credible interval (CrI) differs from the frequentist CI in that its interpretation is directly related to the observed data (probability distribution), whereas the frequentist CI is a characteristic of the statistical method employed. The statistical method merely signifies that CIs, calculated across multiple hypothetical datasets generated by the same underlying process, will encompass the true parameter value in a specific proportion of instances (Hoekstra et al., 2014; Morey et al., 2016). Therefore, CrIs of Bayesian statistics can be more informative about the true effects of the fit model. Third, the degrees of evidence for or against a hypothesis (see Jeffreys, 1961) offer a more comprehensive understanding of the extent to which the observed effects take place compared to the simplistic approach of the "absent–present" effects given by frequentist statistics. *P*-values are not always reliable because they are highly sensitive to sample sizes, failing to provide accurate results when the sample is small (Gelman & Carlin, 2014).

However, Bayesian models do have some disadvantages. For example, they are computationally demanding as it takes more time to fit compared to frequentist models. In addition, more technical knowledge is required to adjust their parameters in a certain way and get the desired results. The researchers also need to be able to understand how these models work in order to interpret the results accordingly. Another important drawback, which is not directly related to the model itself, is its limited popularity in linguistic research (Garcia, 2018). It is true that only a small body of studies in linguistics and even fewer in phonetics have employed Bayesian statistics. In our experience, certain journals have desk-rejected our submitted papers on the grounds of employing Bayesian statistics. The rationale cited was a perceived unfamiliarity among reviewers, hindering their ability to conduct a thorough review and an anticipated challenge for the broader readership in comprehending the content.

In a series of studies examining both linguistic and nonlinguistic research (Haendler et al., 2020; Hong et al., 2013; Norouzian et al., 2019; Samaniego, 2010; Westera, 2021), the comparison between frequentist and Bayesian methods has been a central focus. Norouzian et al. (2019) conducted a comprehensive reevaluation of empirical studies in L2 learning, employing Bayesian hypothesis testing to reassess the outcomes of hundreds of *t*-tests. Their investigation revealed a convergence of inferential conclusions between frequentist and Bayesian approaches. However, a noteworthy observation was made: instances where *p*-values fell between 0.01 and 0.05 in frequentist statistics often amounted to only anecdotal evidence in Bayesian hypothesis testing. This discovery led the authors to question the reliability of *p*-values derived from frequentist statistics as accurate indicators of true signals within the field. Haendler et al. (2020) used both frequentist and Bayesian models to examine the online processing of relative clauses by children with developmental language disorder and children with typical development. The findings implied that the Bayesian model outperformed the frequentist model in accurately predicting distinctions between atypical and typical populations. Specifically, the Bayesian model identified a significant difference for a triple interaction of variables, a result not observed in the frequentist model. This highlights the enhanced sensitivity and discriminative capability of the Bayesian approach in capturing complex interactions within the data, leading to more nuanced and precise predictions in the context of distinguishing between the studied populations.

In this study, we directly compare frequentist and Bayesian statistical models to investigate the accuracy in which they can identify differences between monolingual and bilingual speakers in terms of vowel acoustic features. To our knowledge, this is conducted for the first time in the field of phonetics, aiming to help researchers of the field as well as those conducting research in other areas of linguistics to choose the most appropriate method for their statistical analysis. We employed two distinct modeling approaches using the same set of variables: a linear mixed-effects model and a Bayesian linear regression model. The models had a simple structure to more easily interpret their behavior. Our dataset included the values of the first three formants and the duration of vowels produced by monolingual and bilingual speakers. The dataset was small including 200 observations from 20 speakers. We posit that Bayesian models, given their enhanced sensitivity, are better suited for identifying differences between the two populations, particularly when working with small sample sizes. In contrast, frequentist models may not exhibit the same level of sensitivity and might fail to detect such differences due to the smaller sample size.

## 2 Methodology

### 2.1 Participants

Twenty adult participants, ranging in age from their 20s to 30s, took part in this study. The first group comprised 10 monolingual adult Greek speakers. These individuals were born and raised in Greece and permanently lived in the country. The experimental group consisted of another 10 Albanian-Greek bilinguals (females = 3), who were also born and raised in Greece, being native speakers of both Albanian and Greek. All participants had normal vision and hearing and had never experienced cognitive or language disorders. Before participating, they were briefed on the purpose of the study and their rights, and provided written consent following the principles of the Declaration of Helsinki.

### 2.2 Materials

The study utilized five monosyllabic nonsense words within an /fVf/ context, targeting each of the five Greek vowels. These words were presented within the carrier phrase "Ípes /fVf/ páli" ("You said /fVf/ again"). The chosen context is advantageous as it focuses on a vowel between identical fricative consonants (/f/), simplifying the segmentation process for a more targeted analysis of vowel properties.

### 2.3 Procedure

#### 2.3.1 Production Test

Each participant completed the test individually in quiet rooms. Phrases were presented on paper, and participants were instructed by the researcher to read them as if conversing with a friend. Recordings were made using a professional audio recorder at a 44.1 kHz sampling rate and saved as wav. files with a 24-bit resolution. Each participant produced 10 phrases (5 vowels × 2 repetitions), resulting in a total of 200 phrases across all participants. The phrases were randomized for each participant. Before the test, we ensured that participants could read the sentences appropriately.

### 2.3.2 Segmentation process

The words targeted by the speakers were isolated and processed using Praat (Boersma & Weenink, 2023) with similar procedures as described in Georgiou (2023b) and Georgiou & Dimitriou (2023). Spectrograms and waveforms were visually inspected to pinpoint essential acoustic features, allowing for the measurement of vowel boundaries, including formant frequencies (F1, F2, F3) and vocalic duration. Analysis settings included a 0.025 positive window length, a 50 Hz pre-emphasis, and a spectrogram view range up to 5500 Hz. Frequencies were measured from the end of the burst of the preceding stop consonant /f/ to the onset of the vowel (V). The analysis concluded at the end of the vocalic periodicity, marked by waveform and spectrogram characteristics and the onset of the second consonant /f/. Measurements were taken at the midpoint of the vowel segment to minimize the influence of adjacent sounds. Vowel durations were manually labeled by the researcher, determining the start and end points of each vowel token for duration calculation. To standardize the data, F1, F2, F3, and duration values were normalized using the *z*-score method, expressed by the formula $z = [\chi - \mu] / \sigma$, where $\chi$ represents the raw score, and $\mu$ and $\sigma$ denote the mean and standard deviation of the population, respectively.

### 2.4 Statistics

The first analysis employed four linear mixed-effects models in R (R Core Team, 2023) using the lme4 package (Bates et al., 2023). Normalized *F1, F2, F3*, and *duration* (ms) comprised the dependent variables. *Vowel* (/i e a o u/), *group* (monolingual/bilingual) and their interaction were modelled as fixed factors, while *participant* was modelled as a random factor. We used the emmeans package (Lenth et al., 2023) adjusted with the Bonferroni correction method to investigate differences in the production of certain vowels between the monolingual and bilingual groups.

The second analysis utilized Bayesian regression models in R, employing the brms package (Bürkner, 2023). Four Bayesian models were applied, each featuring normalized *F1, F2, F3* values and duration (ms). *Vowel* (/i e a o u/), *group* (monolingual, bilingual) and their interaction served as fixed factors, while *participant* served as a random factor. Prior selection is crucial for the behavior of the model. The Bayesian models were fit using *weakly* informative priors (Gelman et al., 2017). These comprise a modest expression of prior information (Hamra et al., 2013). They are used in Bayesian statistics to avoid overly influencing the posterior distribution with strong prior beliefs and to allow the data to have a substantial impact on the final results (Lemoine, 2019). In addition, when there is low statistical power arising from small sample sizes, data is regularized (i.e., extreme values are either disallowed or downweighted), leading to a lower chance for Type I error and more accurate results (McElreath, 2015). Thus, due to the absence of preconceived assumptions about data parameter behavior in our dataset and our small dataset, we adopted weakly informative priors. These priors were defined as a student's *t*-distribution with 3 degrees of freedom, a mean of 0, and a standard deviation of 2.5 (Escudero et al., 2022; Georgiou, 2022).

To assess the probability of test hypotheses relative to their alternatives, we employed the Evidence Ratio (ER). According to Jeffreys (1961), an ER > 100 indicates extreme evidence, an ER of 30-100 suggests very strong evidence, and an ER of 10-30 implies strong evidence for a hypothesis. Conversely, an ER of 0.03-0.1 indicates strong evidence, an ER of 0.01-0.10 very strong evidence, and an ER < 0.01 extreme evidence against a hypothesis. In terms of frequentist statistics, an ER > 19 is approximately equal to $\alpha = 0.05$ (Milne & Herff, 2020).

## 3 Results

The results of the mixed-effects models demonstrated that all vowels differed significantly from the Intercept term in terms of F1 and F2, and vowel /i/ differed in terms of F3. In addition, there was a significant effect of group in terms of F1, F2, and duration. The interaction between vowel and group for the monolingual category in each vowel differed from the Intercept in terms of F1. With respect to F2, there was a significant interaction between vowel /e/ and group for the monolingual category. Moreover, a significant interaction between vowel /u/ and group for the monolingual category was found in the duration variable. The results of the linear mixed-effects models are presented in Table 1. The Bonferroni posthoc tests assessed the directional hypothesis that monolinguals will produce higher F1, F2, F3, and duration values than bilinguals. Surprisingly, at $\alpha = 0.05$, it was observed that monolinguals had higher acoustic values compared to bilinguals only for F1 of vowel /a/. The results of the posthoc tests are shown in Table 2.

Table 1: Results of the linear mixed-effects models. Vowel /a/ and bilingual served as Intercept terms.

|    |                            | Estimate | SE    | t-value | p-value |
|----|----------------------------|----------|-------|---------|---------|
| F1 | (Intercept)                | 0.922    | 0.130 | 7.118   | < 0.001 |
|    | vowele                     | -0.915   | 0.108 | -8.480  | < 0.001 |
|    | voweli                     | -2.033   | 0.108 | -18.845 | < 0.001 |
|    | vowelo                     | -1.000   | 0.108 | -9.267  | < 0.001 |
|    | vowelu                     | -1.746   | 0.108 | -16.183 | < 0.001 |
|    | groupmonolingual           | 0.953    | 0.183 | 5.203   | < 0.001 |
|    | vowele:groupmonolingual    | -0.426   | 0.153 | -2.792  | 0.006   |
|    | voweli:groupmonolingual    | -0.729   | 0.153 | -4.775  | < 0.001 |
|    | vowelo:groupmonolingual    | -0.646   | 0.153 | -4.236  | < 0.001 |
|    | vowelu:groupmonolingual    | -0.798   | 0.153 | -5.229  | < 0.001 |
| F2 | (Intercept)                | -0.404   | 0.103 | -3.933  | <0.001  |
|    | vowele                     | 1.099    | 0.131 | 8.398   | <0.001  |
|    | voweli                     | 1.678    | 0.131 | 12.824  | <0.001  |
|    | vowelo                     | -0.423   | 0.131 | -3.234  | 0.001   |
|    | vowelu                     | -0.700   | 0.131 | -5.352  | <0.001  |
|    | groupmonolingual           | 0.333    | 0.145 | 2.293   | 0.024   |
|    | vowele:groupmonolingual    | -0.401   | 0.185 | -2.165  | 0.032   |
|    | voweli:groupmonolingual    | -0.131   | 0.185 | -0.707  | 0.481   |
|    | vowelo:groupmonolingual    | -0.300   | 0.185 | -1.622  | 0.107   |
|    | vowelu:groupmonolingual    | -0.103   | 0.185 | -0.559  | 0.577   |
| F3 | (Intercept)                | -0.530   | 0.264 | -2.006  | 0.053   |

|  |  | Estimate | SE | t-value | p-value |
|---|---|---|---|---|---|
|  | vowele | 0.203 | 0.218 | 0.929 | 0.354 |
|  | voweli | 0.457 | 0.218 | 2.095 | 0.038 |
|  | vowelo | 0.251 | 0.218 | 1.151 | 0.251 |
|  | vowelu | 0.335 | 0.218 | 1.535 | 0.127 |
|  | groupmonolingual | 0.541 | 0.374 | 1.448 | 0.157 |
|  | vowele:groupmonolingual | -0.306 | 0.309 | -0.993 | 0.322 |
|  | voweli:groupmonolingual | -0.112 | 0.309 | -0.362 | 0.718 |
|  | vowelo:groupmonolingual | 0.356 | 0.309 | 1.154 | 0.250 |
|  | vowelu:groupmonolingual | 0.165 | 0.309 | 0.536 | 0.593 |
| Duration | (Intercept) | 0.147 | 0.008 | 17.457 | <0.001 |
|  | vowele | -0.012 | 0.006 | -2.114 | 0.036 |
|  | voweli | -0.029 | 0.006 | -5.195 | <0.001 |
|  | vowelo | -0.014 | 0.006 | -2.451 | 0.015 |
|  | vowelu | -0.023 | 0.006 | -4.147 | <0.001 |
|  | groupmonolingual | 0.039 | 0.012 | 3.241 | 0.003 |
|  | vowele:groupmonolingual | -0.003 | 0.008 | -0.412 | 0.681 |
|  | voweli:groupmonolingual | -0.008 | 0.008 | -1.072 | 0.285 |
|  | vowelo:groupmonolingual | 0.001 | 0.008 | 0.150 | 0.881 |
|  | vowelu:groupmonolingual | -0.019 | 0.008 | -2.471 | 0.014 |

Table 2: Results of the posthoc tests with the Bonferroni correction. The hypotheses tested whether monolinguals > bilinguals in terms of acoustic features for each vowel.

|  | Vowel | Estimate | SE | t-value | p-value |
|---|---|---|---|---|---|
| F1 | i | 0.225 | 0.183 | 1.226 | 1.000 |
|  | e | 0.527 | 0.183 | 2.877 | 0.155 |
|  | a | 0.953 | 0.183 | 5.203 | <0.01 |
|  | o | 0.307 | 0.183 | 1.675 | 1.000 |
|  | u | -0.591 | 0.183 | -3.226 | 1.000 |
| F2 | i | 0.202 | 0.145 | 1.392 | 1.000 |
|  | e | -0.068 | 0.145 | -0.467 | 1.000 |
|  | a | 0.333 | 0.145 | 2.293 | 0.535 |
|  | o | 0.033 | 0.145 | 0.225 | 1.000 |
|  | u | 0.229 | 0.145 | 1.581 | 1.000 |
| F3 | i | 0.429 | 0.374 | 1.149 | 1.000 |
|  | e | 0.235 | 0.374 | 0.628 | 1.000 |
|  | a | 0.541 | 0.374 | 1.448 | 1.000 |
|  | o | 0.897 | 0.374 | 2.401 | 0.496 |
|  | u | 0.814 | 0.374 | 2.177 | 0.824 |
| Duration | i | 0.030 | 0.012 | 2.539 | 0.392 |
|  | e | 0.035 | 0.012 | 2.971 | 0.142 |
|  | a | 0.039 | 0.012 | 3.241 | 0.073 |
|  | o | 0.040 | 0.012 | 3.340 | 0.057 |
|  | u | 0.019 | 0.012 | 1.621 | 1.000 |

The Bayesian CrIs indicate a 95% probability that the true values of the variable estimates (i.e., $\beta$ values) may lie between two positive or two negative plausible values, given the observed data (see Hespanhol et al., 2019). These scenarios would indicate a statistically significant result at a significance level of $\alpha = 0.05$ since both CrIs do not include zero. Looking at Table 3, such differences compared to the Intercept terms exist for all vowels, group, and the interaction between all vowels and the monolingual group category in terms of F1. For F2, differences occurred for all vowels, group, and the interaction between vowel /e/ and the monolingual group category. For F3, only vowel /i/ differed from the Intercept term. Regarding duration, differences were observed for vowels /i u/ and group. The findings of the Bayesian regression models for each acoustic feature are shown in Table 3. To draw direct comparisons with the frequentist models, we only report ER > 19 which is equal to $p < 0.05$. The hypotheses tested whether monolingual speakers exhibited higher acoustic values compared to bilingual speakers. As seen in Table 4, there was evidence that monolinguals had higher values for the F1 of vowels /e a/, the F2 of vowel /a/, the F3 of vowels /o u/, and the duration of vowels /i e a o/. Figure 1 shows the Kernel density estimates of the Bayesian models. Figures 2 and 3 illustrate the formant frequencies and durations of vowels produced by monolingual and bilingual speakers respectively.

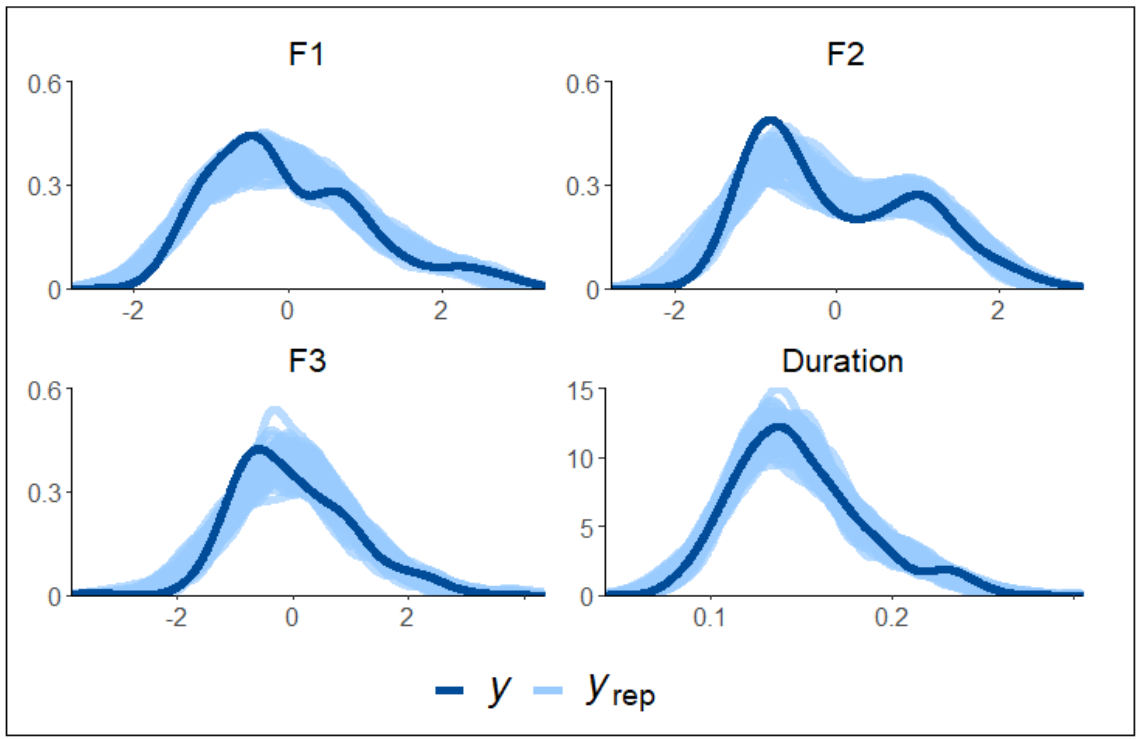

Figure 1: Kernel density estimates of the observed dataset *y* (dark blue curves), with density estimates for 100 simulated datasets *y* rep drawn from the posterior predictive distribution (light blue curves). Each figure illustrates the density of each of the four Bayesian models targeting F1, F2, F3, and duration respectively.

Table 3: Results of the Bayesian regression models

|  |  | β | SE | l-95% CrI | u-95% CrI | Rhat | Bulk ESS | Tail ESS |
|---|---|---|---|---|---|---|---|---|
| F1 | Intercept | 0.92 | 0.14 | 0.64 | 1.18 | 1.00 | 873 | 1530 |
|  | vowele | -0.91 | 0.11 | -1.13 | -0.71 | 1.00 | 1465 | 2042 |
|  | voweli | -2.03 | 0.11 | -2.25 | -1.82 | 1.00 | 1587 | 2219 |
|  | vowelo | -1.00 | 0.11 | -1.21 | -0.78 | 1.00 | 1526 | 2300 |
|  | vowelu | -1.75 | 0.11 | -1.96 | -1.54 | 1.00 | 1601 | 2208 |
|  | groupmonolingual | 0.95 | 0.20 | 0.56 | 1.33 | 1.00 | 860 | 1680 |
|  | vowele:groupmonolingual | -0.42 | 0.15 | -0.71 | -0.11 | 1.00 | 1479 | 2280 |
|  | voweli:groupmonolingual | -0.73 | 0.16 | -1.03 | -0.42 | 1.00 | 1635 | 2171 |
|  | vowelo:groupmonolingual | -0.64 | 0.16 | -0.95 | -0.34 | 1.00 | 1567 | 2337 |
|  | vowelu:groupmonolingual | -0.79 | 0.16 | -1.10 | -0.48 | 1.00 | 1532 | 1915 |
| F2 | Intercept | -0.40 | 0.11 | -0.60 | -0.19 | 1.00 | 1416 | 2002 |
|  | vowele | 1.09 | 0.13 | 0.83 | 1.35 | 1.00 | 1940 | 2536 |
|  | voweli | 1.67 | 0.13 | 1.42 | 1.93 | 1.00 | 1834 | 2403 |
|  | vowelo | -0.43 | 0.13 | -0.68 | -0.17 | 1.00 | 1767 | 2618 |
|  | vowelu | -0.70 | 0.13 | -0.96 | -0.45 | 1.00 | 1806 | 2316 |
|  | groupmonolingual | 0.33 | 0.15 | 0.04 | 0.62 | 1.00 | 1450 | 2021 |
|  | vowele:groupmonolingual | -0.39 | 0.19 | -0.76 | -0.03 | 1.00 | 2165 | 2599 |
|  | voweli:groupmonolingual | -0.12 | 0.18 | -0.49 | 0.24 | 1.00 | 1839 | 2378 |
|  | vowelo:groupmonolingual | -0.29 | 0.19 | -0.65 | 0.08 | 1.00 | 1900 | 2667 |
|  | vowelu:groupmonolingual | -0.10 | 0.19 | -0.46 | 0.26 | 1.00 | 1979 | 2525 |
| F3 | Intercept | -0.50 | 0.27 | -1.02 | 0.05 | 1.00 | 1000 | 1611 |
|  | vowele | 0.19 | 0.21 | -0.23 | 0.60 | 1.00 | 2144 | 2590 |
|  | voweli | 0.44 | 0.21 | 0.03 | 0.84 | 1.00 | 2007 | 2723 |
|  | vowelo | 0.24 | 0.22 | -0.18 | 0.67 | 1.00 | 2051 | 2695 |
|  | vowelu | 0.32 | 0.22 | -0.10 | 0.75 | 1.00 | 2113 | 2524 |
|  | groupmonolingual | 0.50 | 0.39 | -0.26 | 1.24 | 1.00 | 991 | 1443 |
|  | vowele:groupmonolingual | -0.28 | 0.31 | -0.89 | 0.32 | 1.00 | 1974 | 2449 |
|  | voweli:groupmonolingual | -0.09 | 0.30 | -0.67 | 0.50 | 1.00 | 1954 | 2526 |
|  | vowelo:groupmonolingual | 0.37 | 0.31 | -0.23 | 0.97 | 1.00 | 1960 | 2788 |
|  | vowelu:groupmonolingual | 0.17 | 0.30 | -0.41 | 0.78 | 1.00 | 2028 | 2874 |
| Duration | Intercept | 0.15 | 0.01 | 0.13 | 0.17 | 1.01 | 556 | 962 |
|  | vowele | -0.01 | 0.01 | -0.02 | 0.00 | 1.00 | 1556 | 2062 |
|  | voweli | -0.03 | 0.01 | -0.04 | -0.02 | 1.00 | 1377 | 2061 |
|  | vowelo | -0.01 | 0.01 | -0.03 | 0.00 | 1.00 | 1406 | 1870 |
|  | vowelu | -0.02 | 0.01 | -0.03 | -0.01 | 1.00 | 1556 | 2181 |
|  | groupmonolingual | 0.04 | 0.01 | 0.01 | 0.07 | 1.01 | 679 | 1158 |
|  | vowele:groupmonolingual | 0.00 | 0.01 | -0.02 | 0.01 | 1.00 | 1457 | 1792 |
|  | voweli:groupmonolingual | -0.01 | 0.01 | -0.02 | 0.01 | 1.00 | 1318 | 2132 |
|  | vowelo:groupmonolingual | 0.00 | 0.01 | -0.01 | 0.02 | 1.00 | 1433 | 1953 |
|  | vowelu:groupmonolingual | -0.02 | 0.01 | -0.03 | 0.00 | 1.00 | 1604 | 2443 |

Table 4: Results of Bayesian hypothesis testing. The hypotheses tested whether monolinguals > bilinguals in terms of acoustic values for each vowel.

|  | vowel | β | SE | l-CrI | u-CrI | ER | PP |
|---|---|---|---|---|---|---|---|
| F1 | i | 0.22 | 0.19 | -0.10 | 0.54 | 6.89 | 0.87 |
|  | e | 0.52 | 0.20 | 0.21 | 0.84 | 189.48 | 0.99 |
|  | a | 0.95 | 0.20 | 0.62 | 1.27 | Inf | 1.00 |
|  | o | 0.30 | 0.20 | -0.02 | 0.62 | 15.74 | 0.94 |
|  | u | 0.15 | 0.20 | -0.18 | 0.48 | 3.63 | 0.78 |
| F2 | i | 0.21 | 0.15 | -0.03 | 0.46 | 12.61 | 0.93 |
|  | e | -0.06 | 0.15 | -0.31 | 0.18 | 0.48 | 0.33 |
|  | a | 0.33 | 0.15 | 0.08 | 0.57 | 75.92 | 0.99 |
|  | o | 0.03 | 0.15 | -0.21 | 0.28 | 1.46 | 0.59 |
|  | u | 0.23 | 0.15 | -0.01 | 0.48 | 17.10 | 0.94 |
| F3 | i | 0.41 | 0.40 | -0.25 | 1.04 | 5.64 | 0.85 |
|  | e | 0.22 | 0.40 | -0.45 | 0.87 | 2.39 | 0.70 |
|  | a | 0.50 | 0.39 | -0.15 | 1.13 | 8.78 | 0.90 |
|  | o | 0.87 | 0.40 | 0.21 | 1.52 | 55.34 | 0.98 |
|  | u | 0.67 | 0.40 | 0.02 | 1.33 | 21.86 | 0.96 |
| Duration | i | 0.03 | 0.01 | 0.01 | 0.05 | 71.73 | 0.99 |
|  | e | 0.04 | 0.01 | 0.01 | 0.06 | 234.29 | 1.00 |
|  | a | 0.04 | 0.01 | 0.02 | 0.06 | 443.44 | 1.00 |
|  | o | 0.04 | 0.01 | 0.02 | 0.06 | 399.00 | 1.00 |
|  | u | 0.02 | 0.01 | 0.00 | 0.04 | 12.56 | 0.93 |

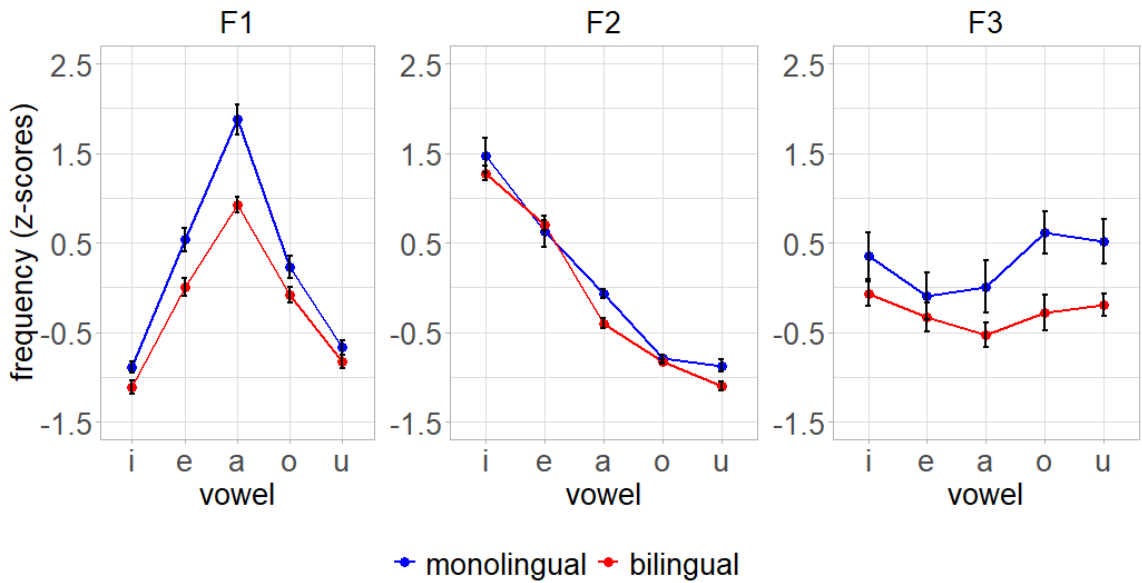

Figure 2: Normalized F1, F2, and F3 values of vowels produced by monolingual and bilingual speakers

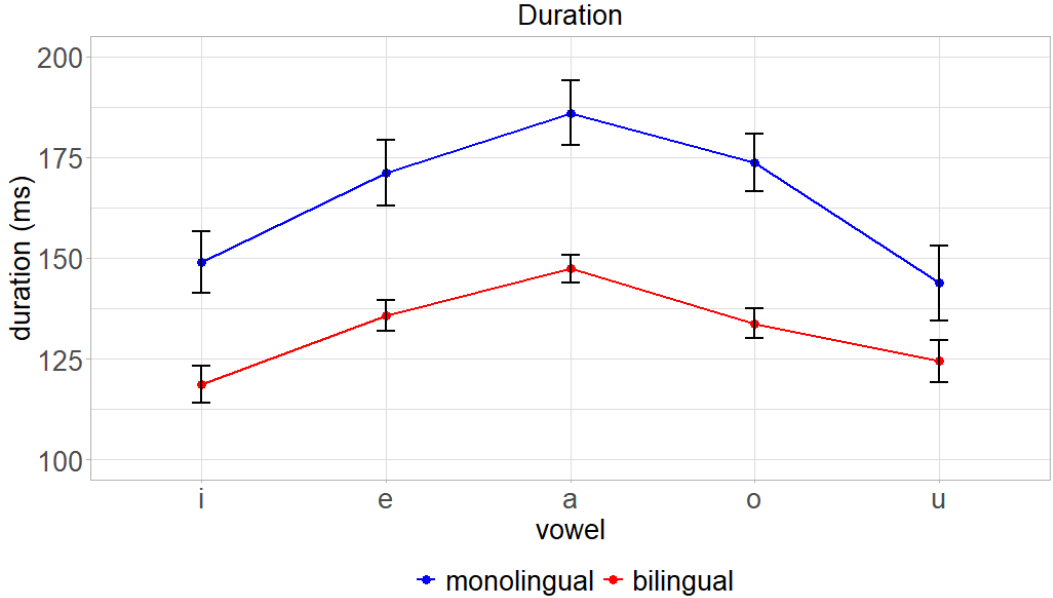

Figure 3: Duration of vowels produced by monolingual and bilingual speakers

**4 Discussion**

This study compared the ability of mixed-effects models and Bayesian regression models to identify differences in the acoustic values of vowels produced by monolingual and bilingual speakers. The goal was to provide researchers of phonetics or language-related fields a basic understanding of how these approaches work and how inferences made using each of them can affect their results and consequently the validity of their studies. We used a small dataset and fit simple models for direct comparison.

Some important conclusions can be drawn about the ability of the frequentist and Bayesian models to identify differences between the monolingual and bilingual populations under examination. The output of the main results of the mixed-effects models is based on the *p*-value. Specifically, $p < 0.05$ provides significant evidence about the existence of an effect, while $p > 0.05$ does not reveal such an effect. By contrast, in Bayesian models, inferences about the existence of significant differences are based on CrIs – if the range of the lowest and highest plausible CrI values do have either a positive or negative sign and do not cross zero, it can be assumed that significant differences occur. In this respect, the behavior of the two approaches as emerges from the current findings is more or less the same. For instance, based on *p*-values, the mixed-effects model indicated significant effects of vowel on the F1 values, with each vowel showing a distinct impact on F1 compared to the reference vowel /a/. It also demonstrated that the monolingual group was associated with a significantly different F1 compared to the bilingual group and that the interaction between all levels of vowels and group for the monolingual level was significant. The same results were obtained through the exploitation of CrIs of the Bayesian model, which did not cross zero.

For the examination of differences between monolinguals and bilinguals for each vowel, we used Bonferroni posthoc tests for the mixed-effects models and hypothesis testing for the Bayesian models. Posthoc analyses involve pairwise comparison tests that assess differences in all possible pairs of means (Jaccard et al., 1984). The Bonferroni correction used in this study is a specific posthoc test which performs independent statistical tests simultaneously on a single data set by dividing the *p*-value by the number of comparisons (e.g., if 10, then $p = α/10$). This is done to avoid Type I errors (Napierala, 2012). By contrast, hypothesis testing uses the Bayes factor, providing the opportunity to the researchers to quantify evidence for two competing hypotheses by comparing how successfully they predict the observed data (van Ravenzwaaij & Wagenmakers, 2022). At the initial stage, which is called likelihood function, the obtained effect size is tested against a limitless number of hypotheses and is benchmarked against all possible hypotheses. The outcome from each benchmarking iteration yields the likelihood of the observed effect based on the corresponding hypothesis. Subsequently, the Bayesian process involves both assigning weights and calculating averages for these probabilities, resulting in a singular value that signifies the probability of the observed effect given the alternative hypothesis. To this end, the Bayes factor is generated. Therefore, a Bayes factor serves as a statistic that conveys the relative evidence in favor of one hypothesis (such as the alternative hypothesis) over another hypothesis (such as the null hypothesis); for a more comprehensive description, see Norouzian et al. (2019). Both testing methods for this study relied on the directional hypothesis $values_{monolingual} > values_{bilingual}$. The two approaches appeared to provide different results. The posthoc test found significant differences ($p < 0.05$) only for F1 of vowel /a/. Conversely, hypothesis testing indicated evidence for the test hypothesis (ER > 19) for F1 of /e a/, F2 of /a/, F3 of /o u/ and duration of /i e a o/. Therefore, hypothesis testing was considerably more sensitive than the posthoc test in identifying differences between the two populations.

The literature suggests that *p*-values present with serious logical and statistical restrictions (e.g., Wagenmakers, 2007), and conclusions drawn solely from these values in the realm of behavioral and social sciences should be approached with caution, as their reliability is called into question (Dienes & Mclatchie, 2017; Kruschke & Liddell, 2018; Rouder et al., 2016). From this study, it is evident that *p*-values derived from mixed-effects models may not be suitable for estimating differences in small datasets containing phonetic data. Such models tend to underestimate the presence of significant differences, a pattern observed not only in our study but also in previous research (e.g., Gallistel, 2009; Rouder et al., 2009). The advantages of Bayesian models over frequentist models in dealing with small datasets compared to frequentist models have been reported in a number of studies (van De Schoot & Depaoli, 2014; van De Schoot et al., 2015; Zhang et al., 2007). Overall, Bayesian analysis proves especially beneficial in scenarios where obtaining large samples is challenging, as observed in atypical populations (refer to Georgiou & Theodorou, 2023), smaller country populations (e.g., Georgiou & Dimitriou, 2023), or within resource-constrained environments such as low-income countries or smaller universities where research funding is limited.

## 5 Conclusions

We discovered that Bayesian models exhibited greater efficiency in detecting differences within a small dataset of phonetic data. This underscores a recommendation for phonetics and linguistics scientists to consider the adoption of such models when working with limited datasets. Subsequent investigations could delve into comparing the performance of frequentist and Bayesian models on larger datasets. Additionally, exploring various Bayesian models with different prior adjustments could offer valuable insights for advancing methodological approaches in phonetics and linguistics research.

**Acknowledgments**

The study is supported by the University of Nicosia Phonetic Lab. We would like to thank the participants of the study. All individual gave their written consent to participate in the study, according to the Declaration of Helsinki.

**Conflicts of interest**

We have no conflicts of interest to disclose.